# Solving the Eikonal equation for compressional and shear waves in anisotropic media using peridynamic differential operator

Ali Can Bekar and Erdogan Madenci[1]
*Department of Aerospace and Mechanical Engr., The University of Arizona, Tucson, AZ USA*

Ehsan Haghighat
*Department of Civil and Environmental Engr.,Massachusetts Institute of Technology, Cambridge, MA, USA*
*Department of Civil Engr.,University of British Columbia, Vancouver, Canada*

Umair bin Waheed
*Department of Geosciences, King Fahd University of Petroleum and Minerals, Dhahran, Saudi Arabia.*

Tariq Alkhalifah
*Physical Sciences and Engr. Division, King Abdullah University of Science and Technology, Thuwal, Saudi Arabia.*

**SUMMARY**

The traveltime of compressional (P) and shear (S) waves have proven essential in many applications of earthquake and exploration seismology. An accurate and efficient traveltime computation for P and S waves is crucial for the success of these applications. However, solutions to the Eikonal equation with a complex phase velocity field in anisotropic media is challenging. The Eikonal equation is a first-order, hyperbolic, nonlinear partial differential equation (PDE) that represents the high-frequency asymptotic approximation of the wave equation. The fast marching and sweeping methods are commonly used due to their efficiency in numercally solving Eikonal equation. However, these methods suffer from numerical inaccuracy in anisotropic media with sharp heterogeneity, irregular surface topography and complex phase velocity fields. This study presents a new method to solving the Eikonal equation by employing the peridynamic differential operator (PDDO). The PDDO provides the nonlocal form of the Eikonal equation by introducing an internal length parameter (horizon) and a weight function with directional nonlocality. The operator is immune to discontinuities in the form sharp changes in field or model variables and invokes the direction of traveltime in a consistent manner. The weight function controls the degree of association among points within the horizon. Solutions are constructed in a consistent manner without upwind assumptions through simple discretization. The capability of this approach is demonstrated by considering different types of Eikonal equations on complex velocity models in anisotropic media. The examples demonstrate its unconditional numerical stability and results compare well with the reference solutions.

[1]Corresponding author. Tel.: +1 520 621 6113.
*E-mail addresses:* acbekar@email.arizona.edu (A. C. Bekar), madenci@email.arizona.edu (E. Madenci), ehsanh@mit.edu (E. Haghighat), umair.waheed@kfupm.edu.sa, (U. b. Waheed), tariq.alkhalifah@kaust.edu.sa (T. Alkhalifah)

# 1 INTRODUCTION

The Eikonal equation is a first-order, hyperbolic, nonlinear partial differential equation (PDE). It represents a high-frequency asymptotic approximation of the wave equation. Despite its origin in optics, the Eikonal equation plays an important role in many practical applications across science and engineering disciplines. In image processing, it is used for distance field computations (Adalsteinsson & Sethian 1995), image denoising (Malladi & Sethian 1996), segmentation (Alvino et al. 2007), and registration (Cao et al. 2004). It is also used to compute geodesic distances in computer graphics for obtaining the shortest paths on discrete and parametric surfaces (Raviv et al. 2011). In robotics, it finds applications for path planning and navigation (Garrido et al. 2016). In semiconductor manufacturing, the Eikonal equation is used for simulating etching, deposition, and lithography (Helmsen et al. 1996). Moreover, in seismology, it is commonly used to perform traveltime tomography (Taillandier et al. 2009), microseismic source localization (Grechka et al. 2015), and Kirchhoff migration (Lambare et al. 2003), etc.

Traditionally, ray-tracing and finite difference methods have been the two most commonly used approaches for solving the Eikonal equation. In the ray tracing approach, a set of linear ordinary differential equations (ODEs), which represents the characteristic solutions of the Eikonal equation, is often employed. The solution to these ODEs offers the parametric description of propagation of wave energy inside the computational domain as a function of time (Bullen 1937). Although the ray tracing method is quite efficient, it exhibits several limitations for practical applications such as the need for traveltime interpolation onto a regular grid for imaging or tomography applications, which becomes difficult when rays diverge due to inhomogeneity in the medium; thus, leading to large spatial gaps. Also, multivalued traveltimes arising from the presence of triplications result in uncertainty in the phase computation. Hence, it becomes difficult to find the path with the minimum traveltime.

In view of these challenges, several techniques have been developed over the years to solve the Eikonal equation numerically (Vidale 1989; Ventura & Ahmad 2015; Sethian 1996). The fast marching method (FMM) and the fast sweeping method (FSM) have emerged as the most commonly accepted techniques due to their efficiency and stability properties. The FMM belongs to the family of algorithms, referred to as the single-pass methods. It has links to Dijkstra's algorithm for the traveling salesman problem (Dijkstra 1959). Dijkstra-like shortest path method is combined with an upwind finite difference scheme for solving the Eikonal equation (Tsitsiklis 1995), which is extended in a level-set context (Sethian 1996) . The main idea of FMM is to track a wavefront evolving in the direction of its normal with a given propagation speed. Unlike the FMM, FSM is a multi-pass algorithm that combines Gauss-Seidel iterations with alternating sweeping direction to solve the Eikonal equation (Zhao 2005). The underlying idea is that the characteristics belonging to the Eikonal equation can be divided into a finite number of pieces and the information propagating along each piece can be accounted for by one of the sweeping directions. Both FMM and FSM were initially developed to solve the Eikonal equation for a medium with an isotropic velocity field. However, the isotropic velocity field is an idealization of the subsurface; the Earth's interior is anisotropic in nature.

Therefore, both FMM and FSM were considered extensively to address anisotropic velocity fields (Grechka et al. 2015; Qin et al. 1992; Luo & Qian 2012) during the last decade. Also, these methods received considerable research interest to attain improvements in the solution accuracy (Popovici & Sethian 2002), incorporate surface topography (Lan & Zhang 2013), and utilize parallel implementation for speed ups in the computation (Zhang & Qian 2004; Detrixhe & Gibou 2016). While these advancements have considerably improved the capabilities of FMM and FSM, they also introduce considerable complexity to the original algorithm by requiring higher-order approximation of the derivatives, particularly in the presence of sharp heterogeneities along with irregular surface topography. Moreover, it is well known that the computational cost of FSM or FMM increases dramatically for anisotropic media compared to the isotropic case.

Recently, there is an increased attention to describe the response of a system through nonlocal integro-differential operators (Silling 2000; Eringen 2002). A recent class of numerical methods, known as Peridynamic Differential Operator (PDDO) (Madenci et al. 2016; Madenci et al. 2017; Madenci et al. 2019), offers a unified approach to solve both local and nonlocal differential equations of arbitrary order using Taylor series expansion. This approach does not require a uniform discretization and is inherently suitable for solving problems with strong heterogeneity. Therefore, it is particularly suitable for geophysical problems where sharp heterogeneities are often the reason for the deficiency encountered with conventional numerical solvers. The PDDO does not involve any spatial derivatives. Therefore, it is immune to discontinuities in the field and model variables and it provides the ability to embed the information on the direction in which information travels in a consistent manner. It enables numerical differentiation through integration; thus, the field equations are valid everywhere regardless of the presence of discontinuities. It simply considers the interaction between neighboring points for the evaluation of derivatives. The weight function controls the degree of association among the points within the horizon. Also, it invokes directional nonlocality based on the knowledge of characteristic directions along which information travels. The solution is achieved through simple discretization without any special treatments. The PDDO has been successfully applied to a variety of partial differential equations and has been used as a data assimilation tool for denoising and data recovery (Madenci et al. 2017; Madenci et al. 2019; Bekar & Madenci 2021).

In order to remove the difficulties faced with FMM and FSM, this study presents a new numerical solution to the Eikonal equation by employing PDDO. The PDDO provides the nonlocal form of the Eikonal equation by introducing an internal length parameter (horizon) and a weight function with directional nonlocality. Therefore, it is extremely suitable for solving the Eikonal equation for traveltimes due compressional (P) and shear (S) waves in sharply heterogeneous velocity fields and in the presence of sharp variations in surface topography in an anisotropic media. Unlike the FSM and FMM, the PDDO solution does not rely on the information from the neighboring grid points. Therefore, it is much easier to parallelize the solver for improvement in its computational efficiency. Furthermore, numerical stability is always ensured and solutions compare well with the reference solutions.

In the next section, we present different forms of the Eikonal equation applicable to seismology studies based on travel time computations. Subsequently, we explain the peridynamic differential operator (PDDO) solution method with a brief discussion on its numerical implementation. Finally, we demonstrate the accuracy of the PD solutions by using (1) the isotropic Marmousi model, (2) the elliptically isotropic Marmousi model, (3) the Hess VTI model, 4) the BP TTI model, and 5) the qSV-wave Eikonal equations previously considered by others using FMM and FSM. We will demonstrate that these solutions are more accurate than the FSM results. While this is the first study in the use of nonlocal PDDO for solving Eikonal equations, it builds the foundation for further studies on the nonlocal considerations in earthquake travel time computations.

## 2 THE EIKONAL EQUATION

The eikonal equation is a time independent first order nonlinear hyperbolic PDE. In geophysics, it describes the traveltime, $T = T(x, z)$, of propagating waves in a 2D medium bounded by $x \in [L_1, L_2]$ and $z \in [W_1, W_2]$. It is subjected to a constraint at the source location given by $T(x_s, z_s) = 0$. The coordinates $x$ and $z$ represent the offset and depth from the source in the domain of interest, respectively. The traveltime depends on the source location, anisotropy and the type of wave propagation.

For P waves in a medium with tilted transverse isotropy (TTI), the Eikonal equation, under the acoustic assumption, is expressed as (Alkhalifah 1998; Alkhalifah 2000; Waheed & Alkhalifah 2017; Waheed et al. 2021)

$$(1 + 2\varepsilon) p_x^2 + p_z^2 \left(1 - R p_x^2\right) = \frac{1}{\upsilon^2} \tag{1}$$

where $p_x$ and $p_z$ represent the horizontal and vertical slowness components. They are defined as

$$p_x = T_{,x} \cos\theta + T_{,z} \sin\theta \tag{2}$$

and

$$p_z = T_{,z} \cos\theta - T_{,x} \sin\theta. \tag{3}$$

Also, the known medium coefficient, $R(x, z)$ is defined as

$$R = \frac{2\eta \upsilon^2 (1 + 2\varepsilon)}{1 + 2\eta} \tag{1}$$

in which $\upsilon = \upsilon(x, z)$ and $\theta = \theta(x, z)$ represent the phase velocity and symmetry axis angle, respectively, and $\varepsilon = \varepsilon(x, z)$ and $\eta = \eta(x, z)$ are referred to as anisotropy parameters. The variable after subscript comma denotes differentiation with respect to.

In a medium with vertical transverse isotropy (VTI) for $\theta = 0$, the Eikonal equation reduces to

$$\left(1 + 2\varepsilon\right) T_{,x}^2 + T_{,z}^2 \left(1 - R T_{,x}^2\right) = \frac{1}{\upsilon^2} \tag{4}$$

This equation can be further simplified to its elliptically isotropic form by setting $\eta = 0$ as

$$\left(1 + 2\varepsilon\right) T_{,x}^2 + T_{,z}^2 = \frac{1}{\upsilon^2} \quad . \tag{5}$$

Furthermore, with $\varepsilon = 0$, it reduces to its simplest form for an isotropic medium as

$$T_{,x}^2 + T_{,z}^2 = \frac{1}{\upsilon^2}. \tag{6}$$

In a medium with TTI, the coupled qP and qSV waves form a quartic surface of slowness given by Han et al. (2017)

$$p_z^4 - B p_z^2 + C = 0 \tag{7}$$

where

$$B = \frac{1}{\upsilon_p^2} + \frac{1}{\upsilon_s^2} - 2\left(1 + \delta + (\varepsilon - \delta)\frac{\upsilon_p^2}{\upsilon_s^2}\right) p_x^2 \tag{8}$$

and

$$C = \left((1 + 2\varepsilon) p_x^2 - \frac{1}{\upsilon_p^2}\right)\left(p_x^2 - \frac{1}{\upsilon_s^2}\right) \tag{9}$$

in which $\upsilon_p = \upsilon_p(x, z)$ and $\upsilon_s = \upsilon_s(x, z)$ represent the phase velocity of the medium in P and S waves in the vertical direction, respectively, and $\varepsilon = \varepsilon(x, z)$ and $\delta = \delta(x, z)$ are refered to as Thomsen parameters. The roots of Eq. (7) represent the qSV- and qP-wave slowness surfaces. The qP-wave travels faster than the qSV- wave; thus, leading to

$$p_z^2 - \frac{B + \sqrt{B^2 - 4C}}{2} = 0 \quad \text{for qSV-wave} \tag{10}$$

and

$$p_z^2 - \frac{B - \sqrt{B^2 - 4C}}{2} = 0 \quad \text{for qP-wave.} \tag{11}$$

## 3  PERIDYNAMIC DIFFERENTIAL OPERATOR

The Peridynamic Differential Operator (PDDO) introduced by Madenci et al. (2016, 2017, 2019) approximates the nonlocal representation of a scalar field $f = f(\mathbf{x})$ and its derivatives at point $\mathbf{x}$ by accounting for the effect of its interactions with the other points, $\mathbf{x}'$, in the domain of interaction, as shown in Fig. 1. It provides differentiation of $N$-th order in $M$ dimensions through integration without a medium smoothness requirement.

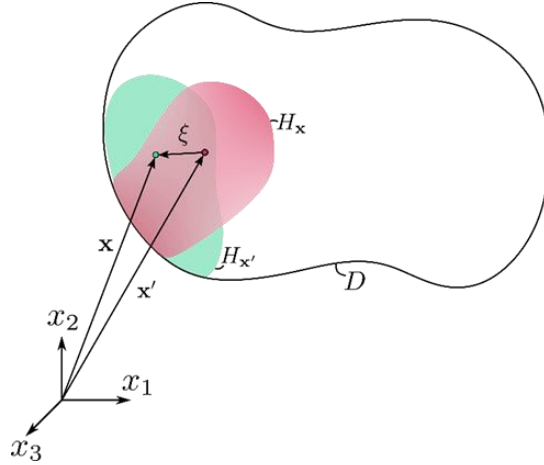

**Figure 1.** Interaction of peridynamic points, $\mathbf{x}$ and $\mathbf{x}'$ with arbitrary family size and shape

The PDDO employs the concept of PD interactions (Madenci et al. 2016; Madenci et al. 2019) and PD functions. The PD functions are orthogonal to each term in the Taylor Series Expansion (TSE). Each point has its own family members in the domain of interaction (family), and occupies an infinitesimally small entity such as volume, area or distance. The points $\mathbf{x}$ and $\mathbf{x}'$ only interact with the other points in their own families, $H_{\mathbf{x}}$ and $H_{\mathbf{x}'}$, respectively. Neither point $\mathbf{x}$ nor $\mathbf{x}'$ is necessarily symmetrically located in their interaction domains. The initial relative position, $\xi$, between the points $\mathbf{x}$ and $\mathbf{x}'$ can be expressed as $\xi = \mathbf{x}' - \mathbf{x}$. This ability permits each point to have its own unique family members with an arbitrary position. Therefore, the size and shape of each family can be different. The degree of interaction between the material points in each family is specified by a non-dimensional weight function, $w(|\xi|)$, which can vary from point to point. Also, it can be modified to invoke directional nonlocality based on the direction of information travel. Thus, the family size and shape are important parameters.

The major difference between the PDDO and other existing local and nonlocal numerical differentiation methods is that the PDDO leads to analytical expressions for arbitrary order derivatives given in integral form for symmetric interaction domains. It provides accurate evaluation of derivatives in the interior as well as the near the boundaries of the domain.

The PDDO for the $N$-th order derivative of a function $f(\mathbf{x})$ with $M$ dimensions can be expressed as (Madenci et al. 2016; Madenci et al. 2019)

$$\frac{\partial^{r_1 + r_2 + \cdots + r_N} f(\mathbf{x})}{\partial x_1^{r_1} \partial x_2^{r_2} \cdots \partial x_M^{r_N}} = \int_{H_{\mathbf{x}}} f(\mathbf{x} + \xi) \, g_N^{r_1 r_2 \cdots r_N}(\xi) d\xi_1 d\xi_2 \cdots d\xi_M \tag{12}$$

in which $r_i$ denotes the order of differentiation with respect to variable $x_i$ with $i = 1, \ldots, M$, and $g_N^{r_1 r_2 \cdots r_N}(\xi)$ are the PD functions explained in detail in a recent book (Madenci et al. 2019).

Since only the first order derivatives of traveltime, $T(x, z)$ appear in the Eikonal equation, its TSE is expressed as

$$T(\mathbf{x} + \xi) = T(\mathbf{x}) + \xi_x \frac{\partial T(\mathbf{x})}{\partial x} + \xi_z \frac{\partial T(\mathbf{x})}{\partial z} + R_1(\mathbf{x}) \tag{13}$$

where $\mathbf{x}^T = \{x, z\}$, $\xi_x$ and $\xi_z$ are the components of the vector, $\boldsymbol{\xi}$ and $R_1(\mathbf{x})$ represents the remainder of the 1$^{st}$ order approximation. Multiplying each term with PD functions, $g_1^{p_1 p_2}(\boldsymbol{\xi})$ and integrating over the domain of interaction (family), $H_{\mathbf{x}}$ result in

$$\int_{H_{\mathbf{x}}} T(\mathbf{x}+\boldsymbol{\xi}) \, g_1^{r_x r_z}(\boldsymbol{\xi}) dV = T(\mathbf{x}) \int_{H_{\mathbf{x}}} g_1^{r_x r_z}(\boldsymbol{\xi}) dH_{\mathbf{x}'} +$$
$$\frac{\partial T(\mathbf{x})}{\partial x} \int_{H_{\mathbf{x}}} \xi_x \, g_1^{r_x r_z}(\boldsymbol{\xi}) dH_{\mathbf{x}'} + \frac{\partial T(\mathbf{x})}{\partial z} \int_{H_{\mathbf{x}}} \xi_z \, g_1^{r_x r_z}(\boldsymbol{\xi}) dH_{\mathbf{x}'} \qquad (14)$$

The PD functions are constructed such that they are orthogonal to each term in the TSE as

$$\int_{H_{\mathbf{x}}} \xi_x^{s_x} \xi_z^{s_z} \, g_1^{r_x r_z}(\boldsymbol{\xi}) dV = \delta_{s_x r_x} \delta_{s_z r_z} \qquad (15)$$

with ( $s_x, s_z, r_x, r_z = 0,1$ ) and $\delta_{s_i r_i}$ as the Kronecker delta symbol. Enforcing the orthogonality conditions in the TSE leads to the nonlocal PD representation of the function itself and its first order derivatives as

$$T^{PD}(\mathbf{x}) = \int_{H_{\mathbf{x}}} T(\mathbf{x}+\boldsymbol{\xi}) \, g_1^{00}(\boldsymbol{\xi}) dH_{\mathbf{x}'} \qquad (16)$$

and

$$\begin{Bmatrix} T_{,x}^{PD} \\ T_{,z}^{PD} \end{Bmatrix} = \int_{H_{\mathbf{x}}} T(\mathbf{x}+\boldsymbol{\xi}) \begin{Bmatrix} g_1^{10}(\boldsymbol{\xi}) \\ g_1^{01}(\boldsymbol{\xi}) \end{Bmatrix} dH_{\mathbf{x}'} \qquad (17)$$

The PD functions can be constructed as a linear combination of polynomial basis functions

$$g_1^{r_x r_z}(\boldsymbol{\xi}) = a_{00}^{r_x r_z} \, w(|\boldsymbol{\xi}|) + a_{x0}^{r_x r_z} \, w(|\boldsymbol{\xi}|)\xi_x + a_{0z}^{r_x r_z} \, w(|\boldsymbol{\xi}|)\xi_z \qquad (18)$$

where $a_{s_x s_z}^{r_x r_z}$ are the unknown coefficients and $w(|\boldsymbol{\xi}|)$ is the weight function. Inserting the PD functions into the orthogonality equation leads to a system of algebraic equations for the determination of the coefficients summarized as

$$\mathbf{Aa} = \mathbf{b} \qquad (19)$$

in which

$$\mathbf{A} = \int\limits_{H_{\mathbf{x}}} w\left(\left|\boldsymbol{\xi}\right|\right) \begin{bmatrix} 1 & \xi_x & \xi_z \\ \xi_x & \xi_x^2 & \xi_x\xi_z \\ \xi_z & \xi_x\xi_z & \xi_z^2 \end{bmatrix} dH_{\mathbf{x}'} \tag{20}$$

$$\mathbf{a} = \begin{bmatrix} a_{00}^{00} & a_{00}^{10} & a_{00}^{01} \\ a_{x0}^{00} & a_{x0}^{10} & a_{x0}^{01} \\ a_{0z}^{00} & a_{0z}^{10} & a_{0z}^{01} \end{bmatrix} \tag{21}$$

and

$$\mathbf{b} = \begin{bmatrix} 1 & 0 & 0 \\ 0 & 1 & 0 \\ 0 & 0 & 1 \end{bmatrix}. \tag{22}$$

Determination of the coefficients, $a_{s_x s_z}^{r_x r_z}$ via $\mathbf{a} = \mathbf{A}^{-1}\mathbf{b}$ establishes the PD functions $g_1^{r_x r_z}(\boldsymbol{\xi})$. The detailed derivations and the associated computer programs can be found in (Madenci et al. 2019).

The PDDO requires spatial integration. Therefore, the integration is performed through a meshless quadrature technique due to its simplicity. The domain is divided into a finite number of cells. Subsequently, the family (interaction domain) of each point is established, and its degree of interaction (weight function) with the family members is specified. The size of the family can be different for each point. Associated with a particular point, the integration is performed by summing the entity of the points within each family.

For a uniform discretization with a grid spacing of $\Delta$, the number of family members is defined by the horizon size of $\delta = m\Delta$. The size of the family members for each point is specified based on the range of $N \leq m \leq N+2$ with $N=1$ representing the highest order spatial derivative appearing in the PDE. The horizon size is specified as $\delta = (N+1)\Delta$ based on its mean value of $(N+1)$. For a fixed value of $m$, the PD representation must converge to the local form as the parameter $\delta$ approaches zero.

For a two dimensional discretization, the domain is discretized as shown in Fig. 2. The interior points can naturally be assigned a symmetric family while the points near the boundary have nonsymmetric families.

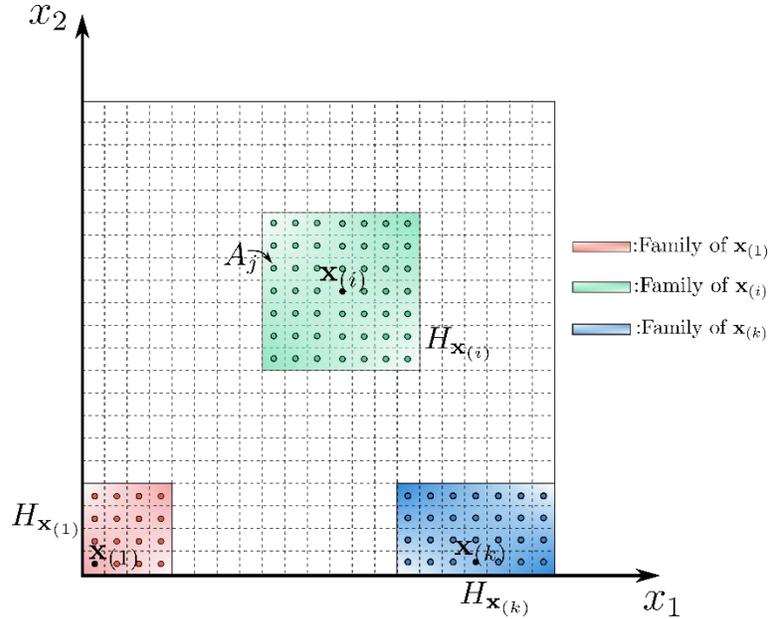

**Figure 2**. PD discretization of a two-dimensional domain and description of a family.

In the discretized form of the equations, the integrations over the family (horizon) in Eqs. (16) and (17) are evaluated as

$$T^{PD}(\mathbf{x}_{(k)}) = \sum_{j=1}^{N_{(k)}} T(\mathbf{x}_{(j)}) \, g_1^{00}(\mathbf{x}_{(j)} - \mathbf{x}_{(k)}) A_{(j)} \quad \text{for} \quad k = 1,...,K \tag{23}$$

and

$$\begin{Bmatrix} T_{,x}^{PD}(\mathbf{x}_{(k)}) \\ T_{,z}^{PD}(\mathbf{x}_{(k)}) \end{Bmatrix} = \sum_{j=1}^{N_{(k)}} T(\mathbf{x}_{(j)}) \begin{Bmatrix} g_1^{10}(\mathbf{x}_{(j)} - \mathbf{x}_{(k)}) \\ g_1^{01}(\mathbf{x}_{(j)} - \mathbf{x}_{(k)}) \end{Bmatrix} A_{(j)} \quad \text{for} \quad k = 1,...,K \tag{24}$$

in which $A_{(j)}$ represents the small incremental area of each PD point, $\mathbf{x}_{(j)}$. The parameter, $N_{(k)}$ represents the number of family members belonging to point, $\mathbf{x}_{(k)}$. Thus, the summation includes all of the contributions from each point, $\mathbf{x}_{(j)}$, within the family of PD point, $\mathbf{x}_{(k)}$. The integration points are always located at the center of its entity in order to achieve unity for the weight in the Gaussian quadrature technique. The parameter, $K$ represents the total number of collocation points in the domain.

For a symmetrically located point in a circular domain of interaction with horizon, $\delta$, the PD functions can be evaluated analytically; thus, leading to

$$T^{PD}(\mathbf{x}) = \frac{1}{2\pi\delta^2} \int_{H_{\mathbf{x}}} w(|\boldsymbol{\xi}|) \, T(\mathbf{x} + \boldsymbol{\xi}) \, dA \tag{25}$$

and

$$\begin{Bmatrix} T_{,x}^{PD} \\ T_{,z}^{PD} \end{Bmatrix} = \frac{2}{\pi\delta^4} \int\limits_{H_x} w(|\xi|) \; T(\mathbf{x}+\xi) \begin{Bmatrix} \xi_x \\ \xi_z \end{Bmatrix} dA \, . \tag{26}$$

The weight function dictates the degree of nonlocality among the points within the family of each point. Any nondimensional weight function is mathematically acceptable; however, in reality, the influence of the weight function should reflect the decrease in the degree of interaction with increasing distances, and the characteristics of the differential equation describing a physical phenomenon.

In order to invoke the direction of travel information during the solution, the PD functions are constructed by employing a weight function with directional nonlocality. The weight function is updated depending on the location of the point inside the family and the gradient direction of the travel time at the point of interest as illustrated in Fig. 3. It is expressed as

$$w\left(|\xi|, \kappa; \delta\right) = \kappa e^{-4(|\xi|/\delta)^2} \tag{27}$$

with

$$\kappa = \begin{cases} 0.1 & \text{if} \quad \xi \cdot \nabla T < 0 \\ 1.0 & \text{if} \quad \xi \cdot \nabla T \geq 0 \end{cases} \tag{28}$$

in which the horizon, $\delta$ defines the extent of domain of interaction, $H_x$ for point $\mathbf{x}$ and the parameter $\kappa$ allows for the information travel with directional nonlocality (upwinding) as illustrated in Fig. 4. This particular form provides a simple way to reflect the effect of upwinding direction and control the degree of interaction between the points, and it ensures numerical stability.

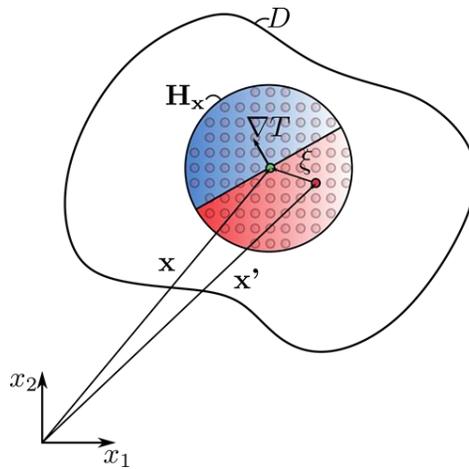

**Figure 3**. Domain of interaction for point $\mathbf{x}$ with directional nonlocality

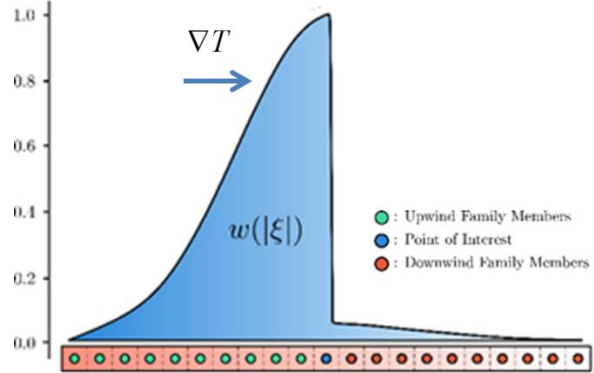

**Figure 4**. Weight function for degree of interaction among the points

## 4 NUMERICAL IMPLEMENTATION

Replacing the local derivatives, $T_{,x}$ and $T_{,z}$ in the Eikonal equation with their PD counterparts, $T_{,x}^{PD}$ and $T_{,z}^{PD}$ leads to its discrete form at each PD point in the domain. Similarly, the constraint condition $T(\mathbf{x}_s) = 0$ is replaced with $T^{PD}(\mathbf{x}_s)$. It is enforced by deleting the corresponding row in the resulting nonlinear algebraic system of equations. The resulting system of equations can be expressed as

$$\mathbf{F}(\mathbf{u}) = \mathbf{0} \tag{29}$$

in which the vector $\mathbf{u}$ contains the PD unknowns, $T(\mathbf{x}_{(k)})$ at each point. These equations can be solved by employing the Newton-Raphson method in an iterative manner. The incremental residual vector, $\Delta\mathbf{F}$ can be obtained from the TSE by considering incremental changes in the unknown vector, $\Delta\mathbf{u}$ as

$$\mathbf{F}(\mathbf{u} + \Delta\mathbf{u}) = \mathbf{F}(\mathbf{u}) + \frac{\partial\mathbf{F}}{\partial\mathbf{u}}\Delta\mathbf{u} + R_1(\mathbf{u}) . \tag{30}$$

Disregarding the remainder terms, $R_1(\mathbf{u})$ leads to the incremental vector, $\Delta\mathbf{F}$ as

$$\Delta\mathbf{F} = \mathbf{F}(\mathbf{u} + \Delta\mathbf{u}) - \mathbf{F}(\mathbf{u}) = \frac{\partial\mathbf{F}}{\partial\mathbf{u}}\Delta\mathbf{u} \tag{31}$$

in which $\Delta\mathbf{u}$ is the unknown incremental vector and $\mathbf{J} = \dfrac{\partial\mathbf{F}}{\partial\mathbf{u}}$ is the Jacobian matrix.

In the numerical implementation, satisfying Eq. (29) by using Eq. (31) in one step is impractical because of the linearization. Hence, the solution is obtained iteratively through a recursive form (fixed point iteration) as

$$\left(\frac{\partial\mathbf{F}}{\partial\mathbf{u}}\right)^{(n)} \Delta\mathbf{u}^{(n+1)} = -\mathbf{F}(\mathbf{u}^n) \tag{32}$$

or

$$\Delta \mathbf{u}^{(n+1)} = -\mathbf{J}^{-1}(\mathbf{u}^{(n)})\mathbf{F}(\mathbf{u}^n) \tag{33}$$

with

$$\mathbf{J}(\mathbf{u}^{(n)}) = \frac{\partial \mathbf{F}(\mathbf{u}^{(n)})}{\partial \mathbf{u}} \tag{34}$$

in which $\mathbf{u}^{(n+1)} = \mathbf{u}^{(n)} + \Delta \mathbf{u}^{(n+1)}$ with $n$ representing the iteration number in the algorithm. The Jacobian matrix is evaluated through automatic differentiation, and the equations are solved by employing the Generalized Minimal RESidual method (GMRES) with ILU preconditioner. The family member construction is achieved by the KD-tree algorithm (Bentley 1975). Depending on the position of family members and the gradient direction of the traveltime, the weight function and the PD functions are updated during each iteration. The convergence of solution is achieved when a specified tolerance is satisfied, i.e., $\|\mathbf{F}(\mathbf{u})\| < e$.

The initial guess for the iterative method is constructed in a consistent manner by modifying the analytical solution (Fomel et al. 2009) for a constant gradient velocity profile as

$$T_{\text{initial}}(\mathbf{x}) = \frac{1}{\sqrt{(\overline{g}_x^2 + \overline{g}_z^2)}} \cosh^{-1}\left(1 + \frac{(\overline{g}_x^2 + \overline{g}_z^2)|\mathbf{x} - \mathbf{x}_s|^2}{2\overline{\upsilon}(\mathbf{x}_s)\overline{\upsilon}(\mathbf{x})}\right) \tag{35}$$

where $\overline{\upsilon}(\mathbf{x})$ represents the velocity field with approximate constant gradients, $\overline{g}_x$ and $\overline{g}_z$. They are approximated as

$$\overline{\upsilon}(\mathbf{x}) = \overline{\upsilon}(x = L_1, z = W_1) + \overline{g}_x x + \overline{g}_z z \tag{36}$$

in which

$$\overline{g}_x \approx \frac{\upsilon(x = L_2, z = (W_2 - W_1)/2) - \upsilon(x = L_1, z = (W_2 - W_1)/2)}{L_2 - L_1} \tag{37a}$$

and

$$\overline{g}_z \approx \frac{\upsilon(x = (L_2 - L_1)/2, z = W_2) - \upsilon(x = (L_2 - L_1)/2, z = W_1)}{W_2 - W_1}. \tag{37b}$$

If the velocity field is coarse, this initial case is used after smoothing by using a Gaussian filter. With the initial guess from Eq. (24), the solution to the smoothed field is used as an initial guess for the coarse velocity field.

The boundedness of the solution **u** depends on the behavior of $-\mathbf{J}^{-1}$. Therefore, the method is stable if the real parts of all eigenvalues of $-\mathbf{J}^{-1}$ are negative. Otherwise, it is unstable. It is worth noting that the real parts of the eigenvalues of $-\mathbf{J}^{-1}$ and $-\mathbf{J}$ have the same signs. Therefore, examining the eigen spectrum of $-\mathbf{J}$ is sufficient. Fig. 5 shows the real and imaginary parts of the eigenvalues of $-\mathbf{J}$ corresponding to the weight function with and without directional nonlocality. The Jacobian, $-\mathbf{J}$ corresponds to the %2 randomly perturbed analytical solution of the isotropic Eikonal equation for the velocity field, $\upsilon(\mathbf{x})$ constant gradients $g_x = 0$ and $g_z = 0.5 s^{-1}$ of the form (Fomel et al. 2009)

$$T(\mathbf{x}) = 2\cosh^{-1}\left(1 + \frac{\left|\mathbf{x} - (1.0,1.0)\right|^2}{20\upsilon(\mathbf{x})}\right) \tag{38}$$

in which the velocity at source $\upsilon(1.0,1.0) = 2.5 \text{ km/s}$. The weight function with directional nonlocality results in eigenvalues with only negative real parts. However, the symmetric weight function without directional nonlocality results in eigenvalues with both positive and negative real parts. This ascertains the stability and necessity of the modification of the symmetric weight function using the gradient of the travel times. Also, the weight function with directional nonlocality ensures the numerical stability of the solution procedure.

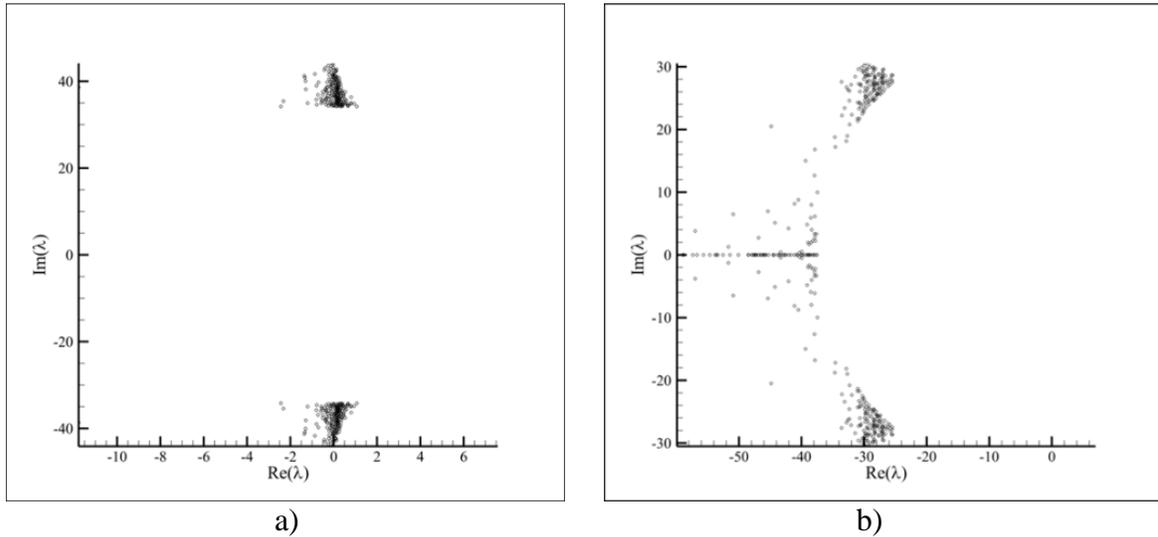

a)                                        b)

**Figure 5**. Real and imaginary parts of the eigenvalues of $-\mathbf{J}$ corresponding weight function: a) without directional nonlocality (symmetric), and b) with directional nonlocality.

## 5  NUMERICAL RESULTS

In order to demonstrate the robustness and simplicity of the present approach, the numerical results concern traveltime solutions due to both compressional (qP) and shear (qSV) waves. The PD solutions are obtained by applying the same procedure and discretization parameters discussed above without any special treatment. The horizon size is specified as $\delta = 2\Delta$ and the initial guess is constructed in a consistent manner. The PD qP traveltime solutions are constructed for (1) Isotropic Marmousi, (2) Elliptically isotropic Marmousi, (3) Hess VTI, 4) BP TTI models due to the availability of FSM solutions for comparison. For a meaningful comparison, the PD discretization is set to the same grid spacing as that of FSM. As shown in the subsequent sections, the PD predictions are closer to the reference solution (second-order FSM with fine mesh) than those of FSM. Without any special treatment, the PD qSV solution is also readily constructed in an anisotropic medium. However, the predictions are compared with the wave field extrapolation using the low-rank approximation. The PD predictions recover the low-rank solution wave front corresponding to the main branch of the wave field. The same procedure applies to the solution of qP traveltime using Eq. (11).

## 5.1 Isotropic Eikonal equation

Replacing the local derivatives with their PD counterparts from Eq. (24) in the Eikonal equation, Eq. (6) leads to its discrete form as

$$\left(\sum_{j=1}^{N_{(k)}} T_{(k)(j)} g_1^{10}(\xi_{1(k)(j)}, \xi_{2(k)(j)}) A_{(j)}\right)^2 + \left(\sum_{j=1}^{N_{(k)}} T_{(k)(j)} g_1^{01}(\xi_{1(k)(j)}, \xi_{2(k)(j)}) A_{(j)}\right)^2 = \frac{1}{\upsilon^2(x_{(k)}, z_{(k)})} \quad \text{for } k = 1, K \quad (39)$$

with $x \in [0, 3.0 \text{ km}]$ and $z \in [0, 3.0 \text{ km}]$. Considering the coarse Marmousi velocity model with a source at $\mathbf{x}_s = (1.5 \text{km}, 1.5 \text{km})$ shown in Fig. 6., the domain is discretized with $K = 101 \times 101$ points. The corresponding grid spacing is $\Delta = 30 \text{ m}$. The initial guess for the smoothed velocity field is constructed with the average velocity gradients $\bar{g}_x = 0$ and $\bar{g}_z = 1.5 s^{-1}$. With $\bar{\upsilon}(\mathbf{x}_s) = 3.75 \text{km/s}$, the initial guess for traveltime becomes

$$T_{init}(\mathbf{x}) = \frac{1}{1.5} \cosh^{-1}\left(1 + \frac{1.5^2 |\mathbf{x} - (1.0, 1.0)|^2}{7.5 \bar{\upsilon}(\mathbf{x})}\right). \quad (40)$$

The solution of the smoothed velocity field becomes the initial guess for the coarse velocity field. Fig. 7 shows the comparison of the PD solution with those of reference solution and FSM. It is clear that the PD solution is closer to the reference solution than that of FSM. The PD solution is achieved when $\|\mathbf{F}(\mathbf{u})\| < 9 \times 10^{-3}$ and the error comparison against the reference solution is shown in Fig. 8. The error is relatively uniform and smaller than that of FSM.

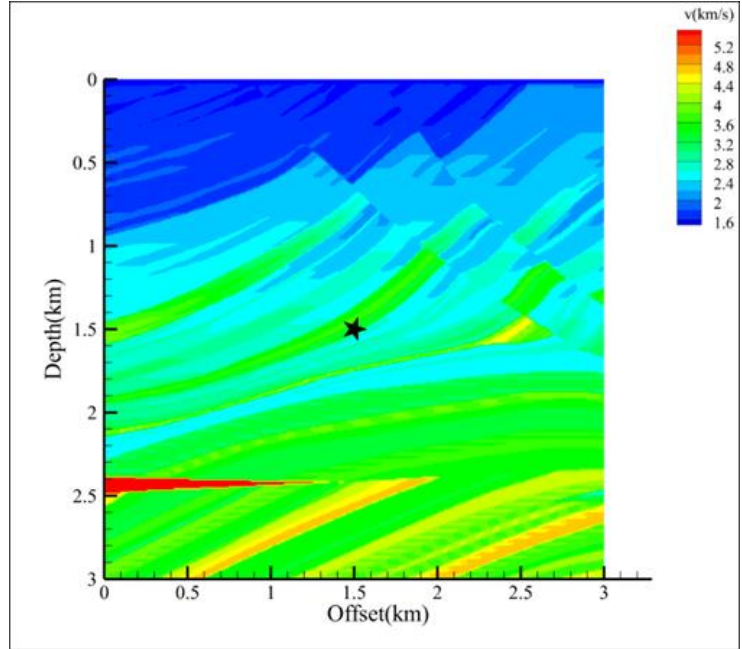

**Figure 6**. Marmousi velocity field in the domain with a source at the center

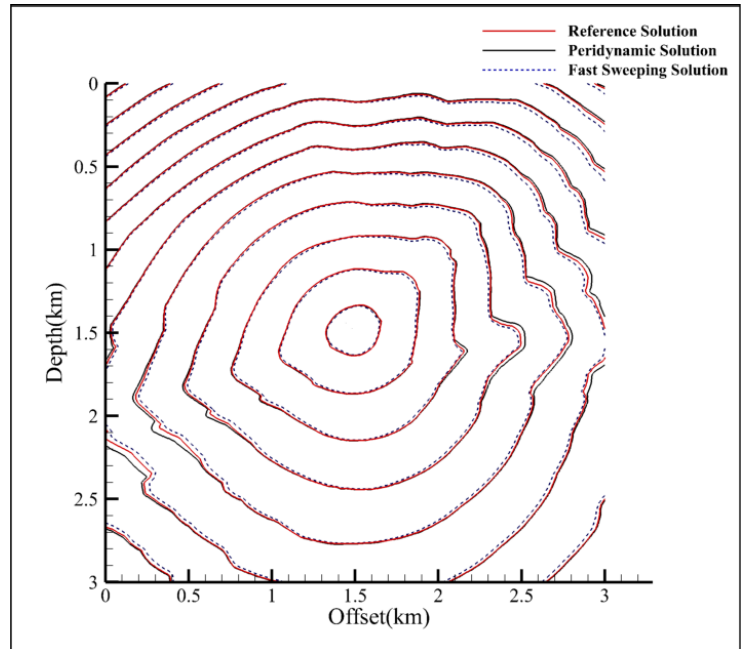

**Figure 7**. Comparison of PD traveltime predictions with the reference and FSM - Isotropic Eikonal equation

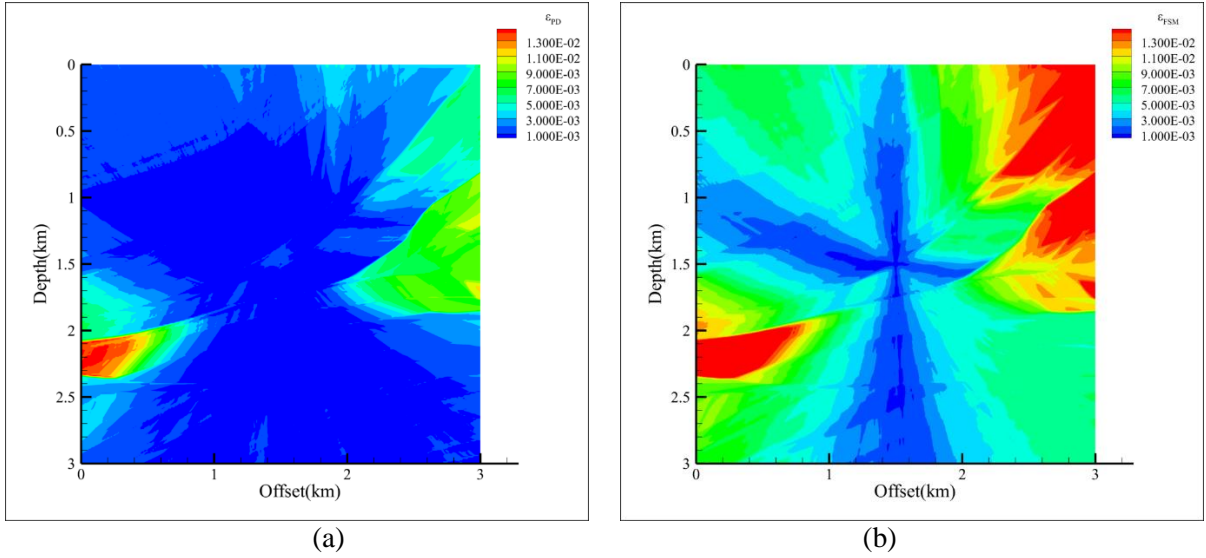

**Figure 8**. Absolute error against reference solution: a) PD, and ) FSM - Isotropic Eikonal equation

## 5.2 Elliptic Eikonal equation

With the PD representation of the local derivatives from Eq. (24), the discrete form of the Elliptic Eikonal equation, Eq. (5) can be written as

$$
\left(1+2\varepsilon(x_{(k)},z_{(k)})\right)\left(\sum_{j=1}^{N_{(k)}}T_{(k)(j)}g_1^{10}(\xi_{1(k)(j)},\xi_{2(k)(j)})A_{(j)}\right)^2 \\
+\left(\sum_{j=1}^{N_{(k)}}T_{(k)(j)}g_1^{01}(\xi_{1(k)(j)},\xi_{2(k)(j)})A_{(j)}\right)^2 = \frac{1}{\upsilon^2(x_{(k)},z_{(k)})} \qquad \text{for } k=1,K \qquad (41)
$$

with $x \in [0, 2.9875\text{km}]$ and $z \in [0, 2.9875\text{km}]$. The velocity field shown in Fig. 6 has a source at $\mathbf{x}_s = (1.5\text{km}, 2.5\text{km})$. The variation of anisotropy parameter, $\varepsilon(x,z)$, shown in Fig. 9 has abrupt changes in the domain. The domain is discretized with $K = 240 \times 240$ points. The corresponding grid spacing is $\Delta = 12.5$ m. The initial guess for the smoothed velocity field is constructed with the average velocity gradients $\overline{g}_x = 0$ and $\overline{g}_z = 1.5 s^{-1}$. With $\overline{\upsilon}(\mathbf{x}_s) = 5.25\text{km/s}$, the initial guess for traveltime becomes

$$
T_{init}(\mathbf{x}) = \frac{1}{1.5}\cosh^{-1}\left(1+\frac{1.5^2\left|\mathbf{x}-(1.5,2.5)\right|^2}{10.5\overline{\upsilon}(\mathbf{x})}\right). \qquad (42)
$$

The solution to the smoothed velocity field becomes the initial gues for the coarse velocity field.

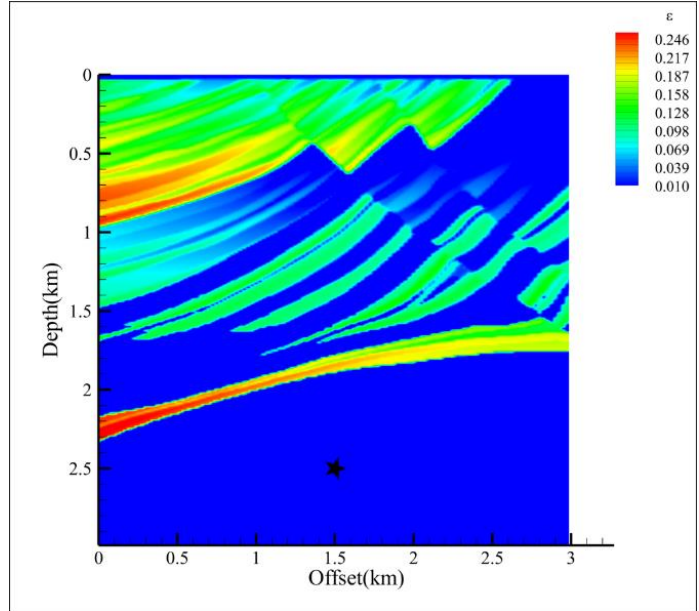

**Figure 9.** Variation of anisotropic parameter $\varepsilon(x, z)$ - Elliptic Eikonal equation

Fig. 10 shows the comparison of the PD solution with those of reference solution and FSM. The PD solution is achieved when $\|\mathbf{F(u)}\| < 2.3 \times 10^{-3}$ and the error comparison against the reference solution is shown in Fig. 11. It obvious that PD performs better than that of FSM.

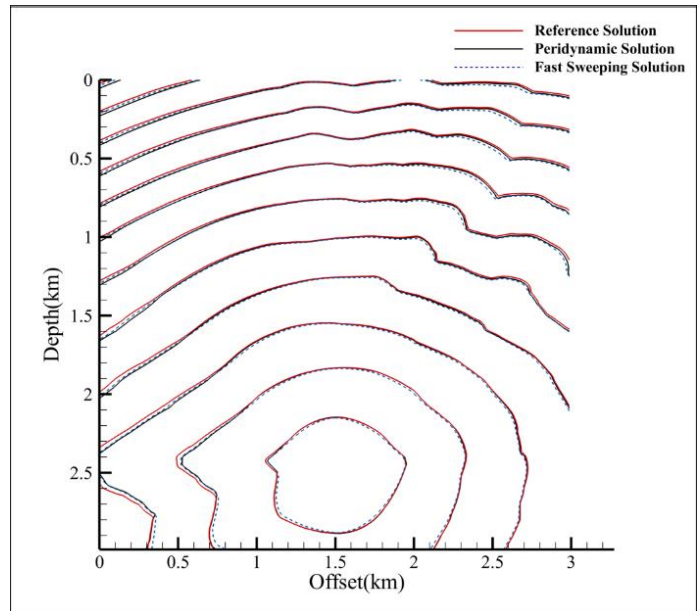

**Figure 10**. Comparison of PD traveltime predictions with the reference and FSM - Elliptic Eikonal equation

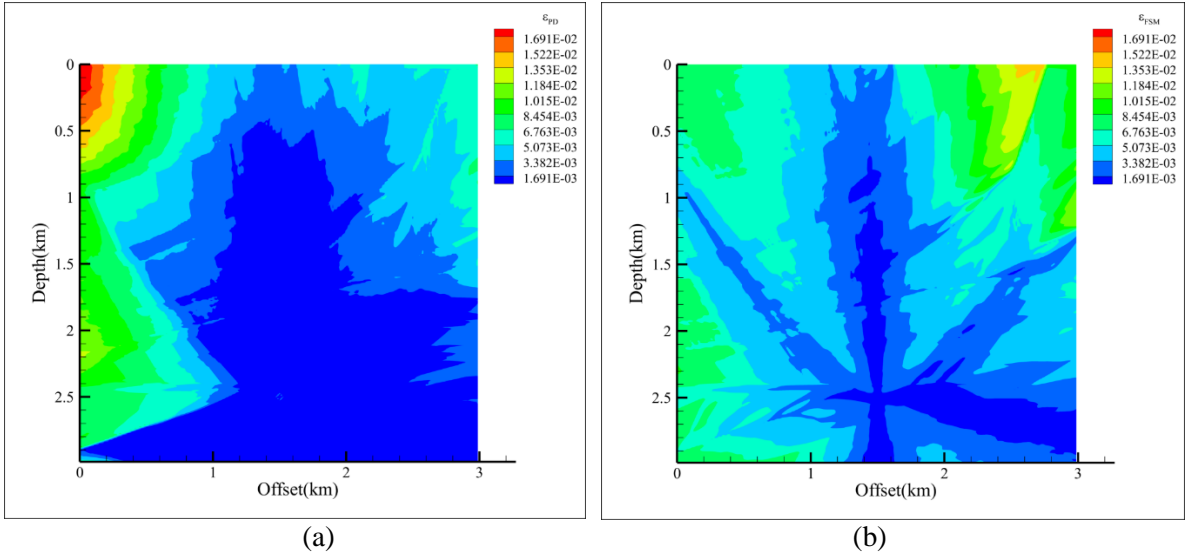

**Figure 11**. Absolute error against reference solution: a) PD, and b) FSM - Elliptic Eikonal equation

### 5.3 Hess VTI Model

Substituting for the local derivatives from Eq. (24) in Eq. (4) results in the discrete form of the VTI Eikonal equation as

$$
\left(1 + 2\varepsilon(x_{(k)}, z_{(k)})\right)\left(\sum_{j=1}^{N_{(k)}} T_{(k)(j)} g_1^{10}(\xi_{1(k)(j)}, \xi_{2(k)(j)}) A_{(j)}\right)^2 +
$$
$$
\left(\sum_{j=1}^{N_{(k)}} T_{(k)(j)} g_1^{01}(\xi_{1(k)(j)}, \xi_{2(k)(j)}) A_{(j)}\right)^2 \times \qquad\qquad \text{for} \quad k = 1, K \qquad (43)
$$
$$
\left(1 - R(x_{(k)}, z_{(k)})\left(\sum_{j=1}^{N_{(k)}} T_{(k)(j)} g_1^{10}(\xi_{1(k)(j)}, \xi_{2(k)(j)}) A_{(j)}\right)^2\right) = \frac{1}{\upsilon^2(x_{(k)}, z_{(k)})}
$$

with

$$
R(x_{(k)}, z_{(k)}) = \frac{2\eta(x_{(k)}, z_{(k)})\upsilon^2(x_{(k)}, z_{(k)})\left(1 + 2\varepsilon(x_{(k)}, z_{(k)})\right)}{1 + 2\eta(x_{(k)}, z_{(k)})} \qquad\qquad (44)
$$

with $x \in [0, 21.69\text{km}]$ and $z \in [0, 8.97\text{km}]$. The velocity field shown in Fig. 12 has a source at $\mathbf{x}_s = (12\text{km}, 3\text{km})$. The variation of anisotropy parameters, $\varepsilon(x, z)$ and $\eta(x, z)$ are shown in Figs. 13 and 14, respectively. The domain is discretized with $K = 724 \times 300$ points. The corresponding grid spacing is $\Delta = 30\,\text{m}$. The initial guess for the traveltime field is constructed with the average velocity gradients $\bar{g}_x = 0$ and $\bar{g}_z = 0.3s^{-1}$. With $\bar{\upsilon}(\mathbf{x}_s) = 2.4\text{km/s}$, the initial guess for traveltime becomes

$$T_{init}(\mathbf{x}) = \frac{1}{0.3}\cosh^{-1}\left(1 + \frac{0.3^2\left|\mathbf{x} - (12.0, 3.0)\right|^2}{4.8\bar{\upsilon}(\mathbf{x})}\right). \tag{45}$$

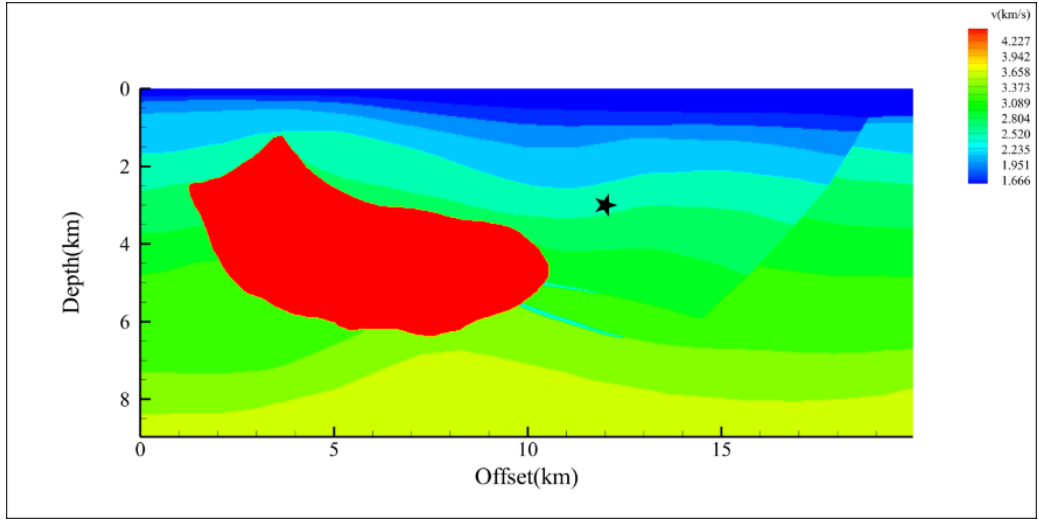

**Figure 12.** Velocity field $\upsilon(x,z)$ in the domain with a source at $\mathbf{x}_s = (12.0, 3.0)$ - Hess model

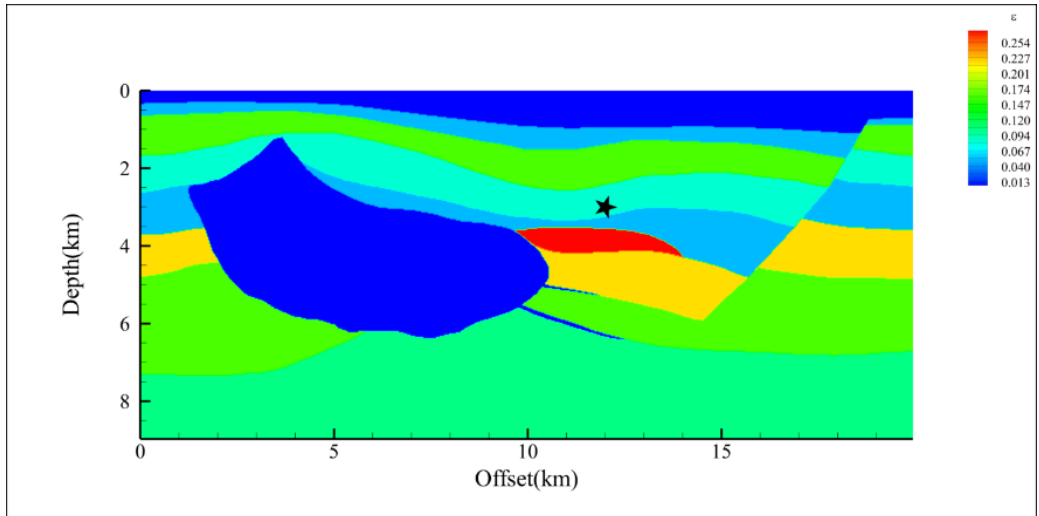

**Figure 13.** Variation of anisotropy parameter $\varepsilon(x,z)$ in the domain - Hess model

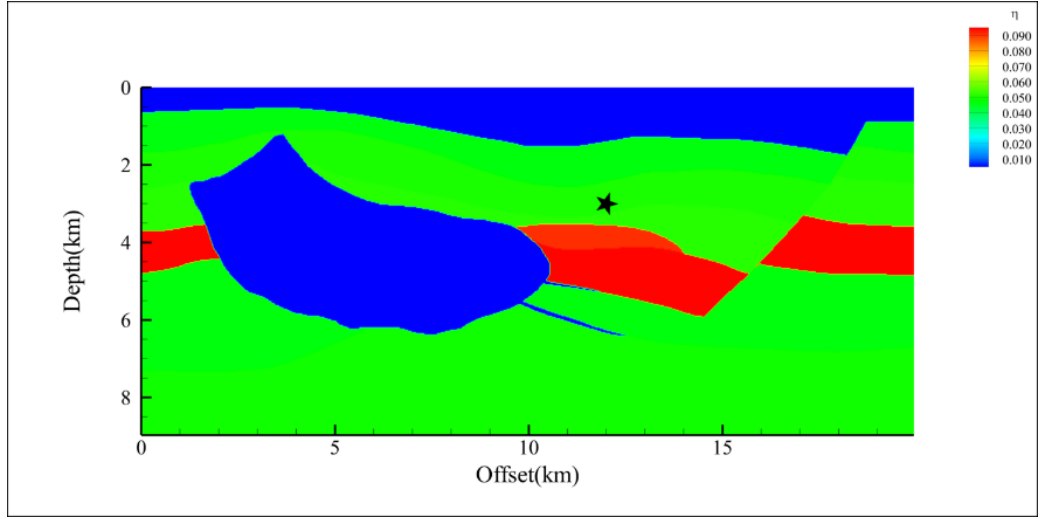

**Figure 14.** Variation of anisotropy parameter $\eta(x, z)$ in the domain - Hess model

Fig. 15 shows the comparison of the PD solution with those of reference solution and FSM. The PD solution is closer to the reference solution than that of FSM. The PD solution is achieved when $\|\mathbf{F(u)}\| < 1.554 \times 10^{-3}$ and the error comparison against the reference solution is shown in Fig. 16. The error is relatively uniform and smaller than that of FSM as shown in Fig. 17.

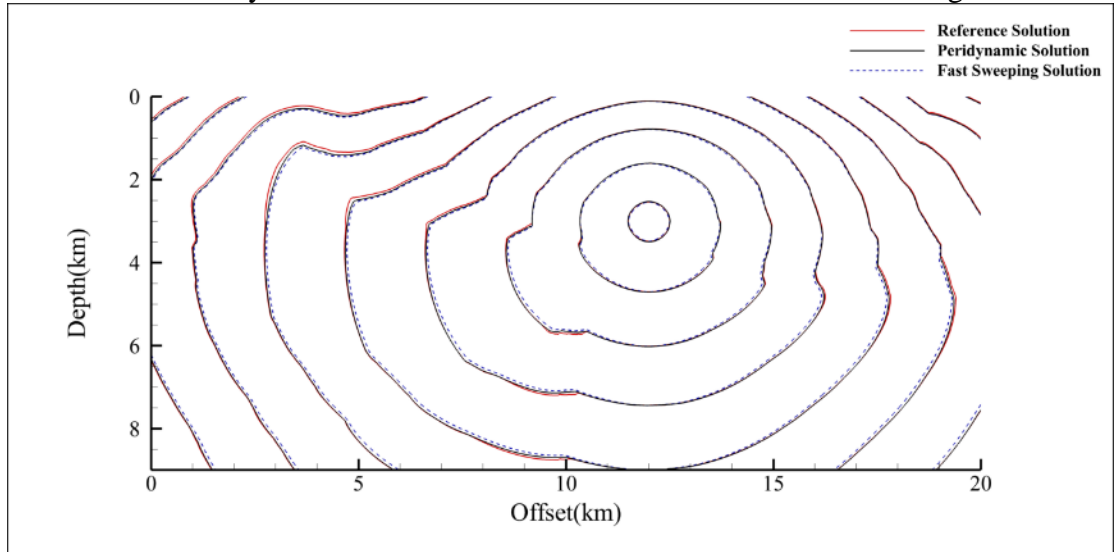

**Figure 15.** Comparison of PD traveltime predictions with the reference and FSM – Hess VTI model

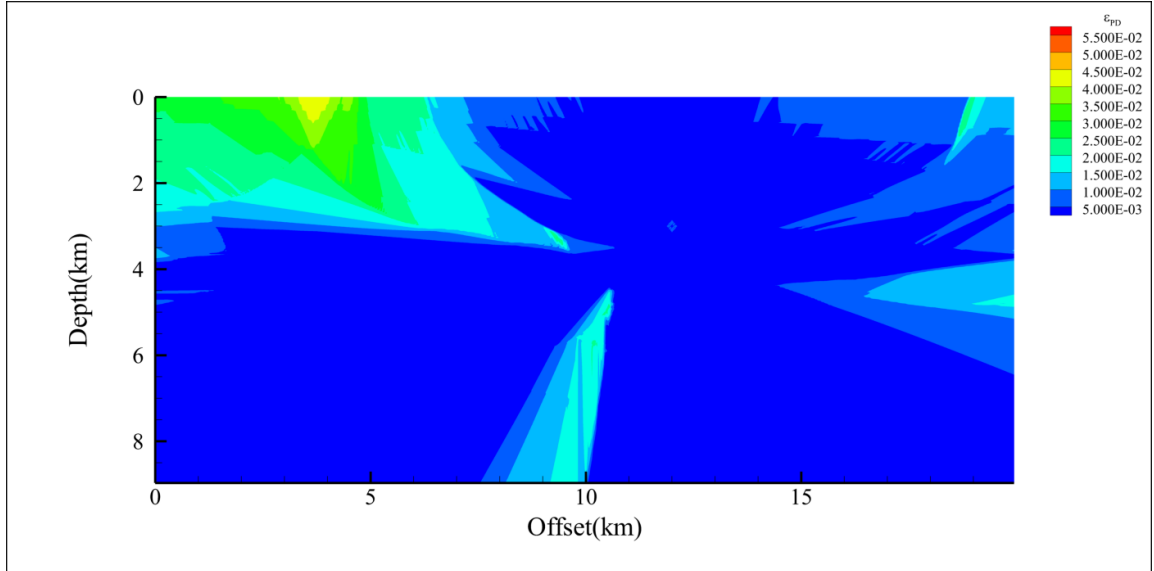

**Figure 16**. Absolute error measure in PD solution against reference solution – Hess VTI model

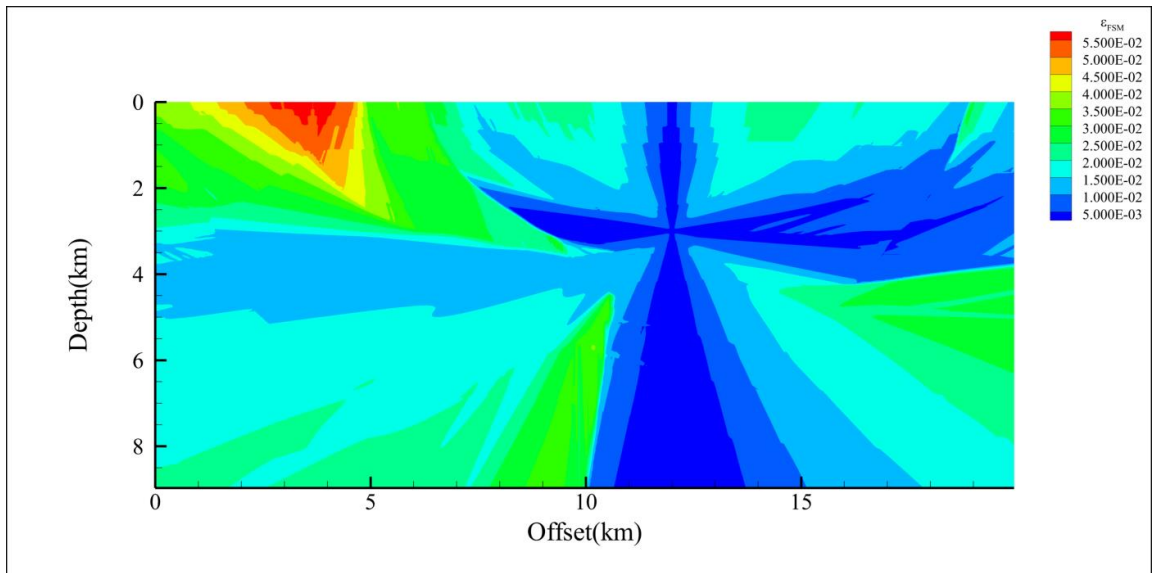

**Figure 17**. Absolute error measure in FSM against reference solution – Hess VTI model

## 5.4 BP TTI Model

Substituting for the local derivatives from Eq. (24) in Eq. (4) results in the discrete form of the VTI Eikonal equation as

$$\left(1 + 2\varepsilon(x_{(k)}, z_{(k)})\right) \left( \left( \sum_{j=1}^{N_{(k)}} T_{(k)(j)} g_1^{10}(\xi_{1(k)(j)}, \xi_{2(k)(j)}) A_{(j)} \right) \cos\left(\theta(x_{(k)}, z_{(k)})\right) + \left( \sum_{j=1}^{N_{(k)}} T_{(k)(j)} g_1^{01}(\xi_{1(k)(j)}, \xi_{2(k)(j)}) A_{(j)} \right) \sin\left(\theta(x_{(k)}, z_{(k)})\right) \right)^2$$

$$+ \left( \left( \sum_{j=1}^{N_{(k)}} T_{(k)(j)} g_1^{01}(\xi_{1(k)(j)}, \xi_{2(k)(j)}) A_{(j)} \right) \cos\left(\theta(x_{(k)}, z_{(k)})\right) - \left( \sum_{j=1}^{N_{(k)}} T_{(k)(j)} g_1^{10}(\xi_{1(k)(j)}, \xi_{2(k)(j)}) A_{(j)} \right) \sin\left(\theta(x_{(k)}, z_{(k)})\right) \right)^2$$

$$\times \left( 1 - R(x_{(k)}, z_{(k)}) \left( \left( \sum_{j=1}^{N_{(k)}} T_{(k)(j)} g_1^{10}(\xi_{1(k)(j)}, \xi_{2(k)(j)}) A_{(j)} \right) \cos\left(\theta(x_{(k)}, z_{(k)})\right) + \left( \sum_{j=1}^{N_{(k)}} T_{(k)(j)} g_1^{01}(\xi_{1(k)(j)}, \xi_{2(k)(j)}) A_{(j)} \right) \sin\left(\theta(x_{(k)}, z_{(k)})\right) \right)^2 \right) = \frac{1}{\upsilon^2(x_{(k)}, z_{(k)})}$$

$$\text{for } k = 1, K \qquad (46)$$

with $x \in [0, 10\text{km}]$ and $z \in [0, 10\text{km}]$. The velocity field shown in Fig. 18 has a source at $\mathbf{x}_s = (5\text{km}, 0.5\text{km})$. The variation of anisotropy parameters, $\varepsilon(x, z)$ and $\eta(x, z)$ and tilt angle, $\theta(x, z)$ are shown in Figs. 19, 20 and 21, respectively. Although the topography at the surface is irregular, the domain is discretized with $K = 161 \times 161$ points. The corresponding uniform grid spacing is $\Delta = 62.5$ m. The initial guess for the traveltime field is constructed with the average velocity gradients $\bar{g}_x = 0$ and $\bar{g}_z = 0.35 s^{-1}$. With $\bar{\upsilon}(\mathbf{x}_s) = 1.175 \text{km/s}$, the initial guess for traveltime becomes

$$T_{init}(\mathbf{x}) = \frac{1}{0.35} \cosh^{-1}\left( 1 + \frac{0.35^2 \left| \mathbf{x} - (5,5) \right|^2}{2.35 \bar{\upsilon}(\mathbf{x})} \right). \qquad (47)$$

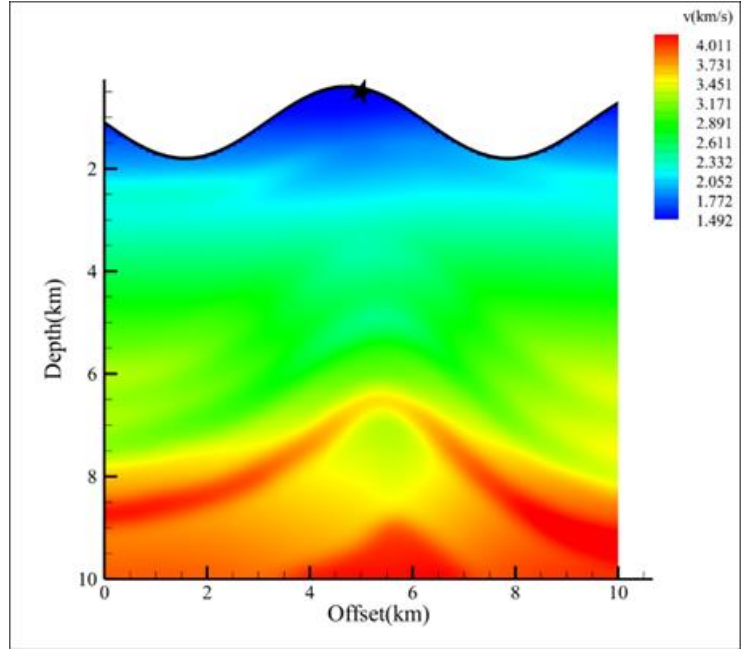

**Figure 18**. Velocity field, $\upsilon(x,z)$ in the domain with a source at $\mathbf{x}_s = (5.0, 0.5)$ - BP TTI model

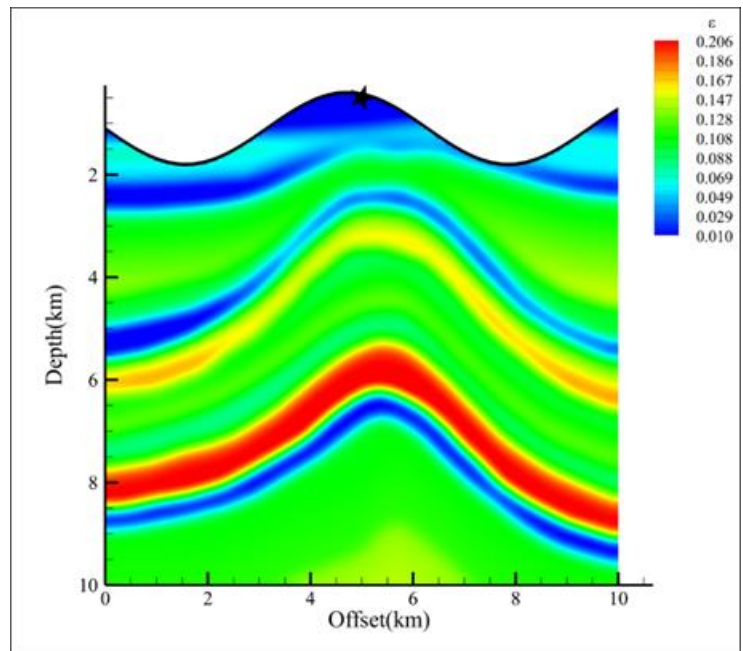

**Figure 19**. Variation of anisotropy parameter $\varepsilon(x,z)$ in the domain - BP TTI model

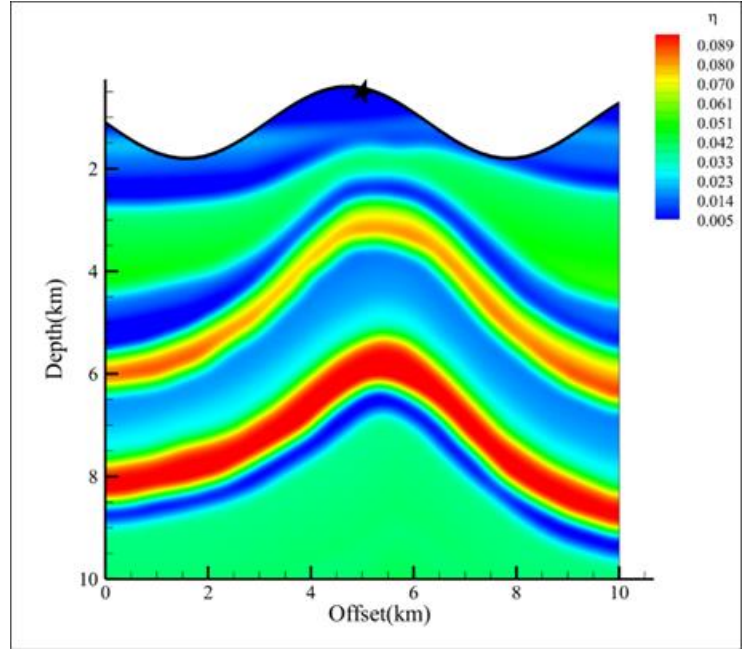

**Figure 20**. Variation of anisotropy parameter $\eta(x, z)$ in the domain - BP TTI model

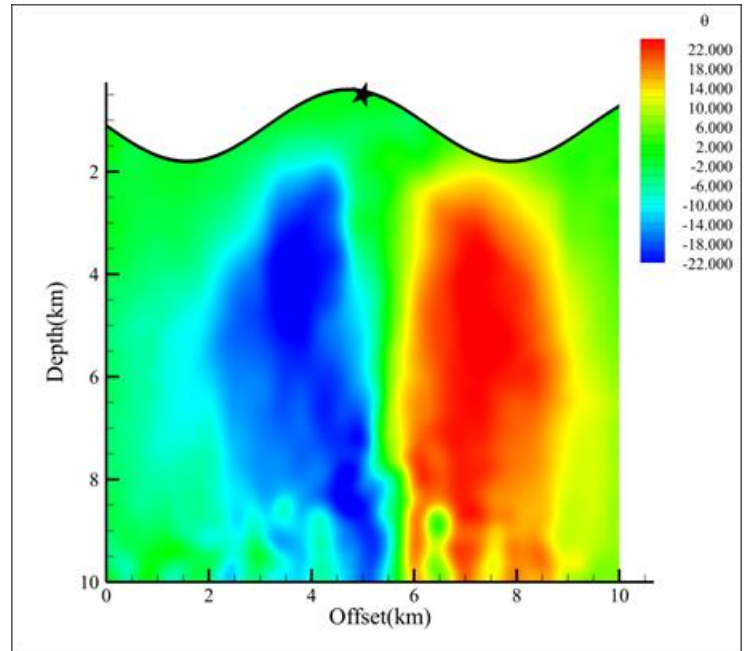

**Figure 21**. Variation of tilte angle $\theta(x, z)$ in the domain - BP TTI model

Fig. 22 shows the comparison of the PD solution with those of reference solution and FSM. The PD solution is closer to the reference solution than that of FSM. The PD solution is achieved when $\|\mathbf{F}(\mathbf{u})\| < 3.17 \times 10^{-3}$ and the error comparison against the reference solution is shown in Fig. 23. The error is relatively uniform and smaller than that of FSM as shown in Fig. 24.

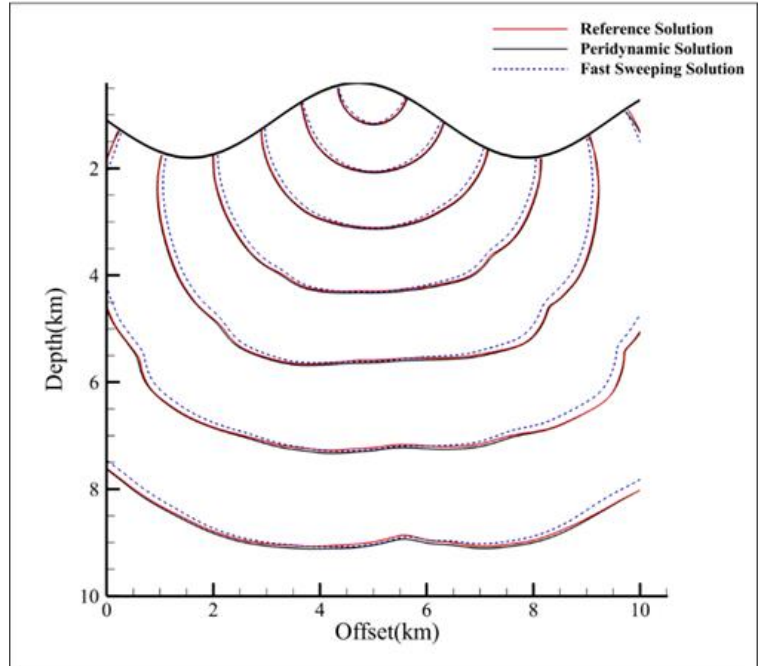

**Figure 22.** Comparison of PD traveltime predictions with the reference and FSM – BP TTI model

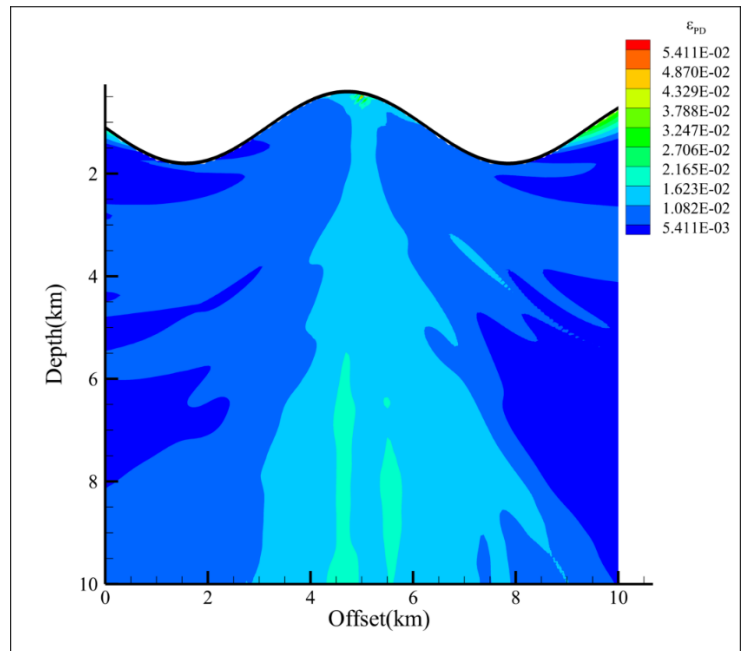

**Figure 23.** Absolute error measure in PD solution against reference solution – BP TTI model

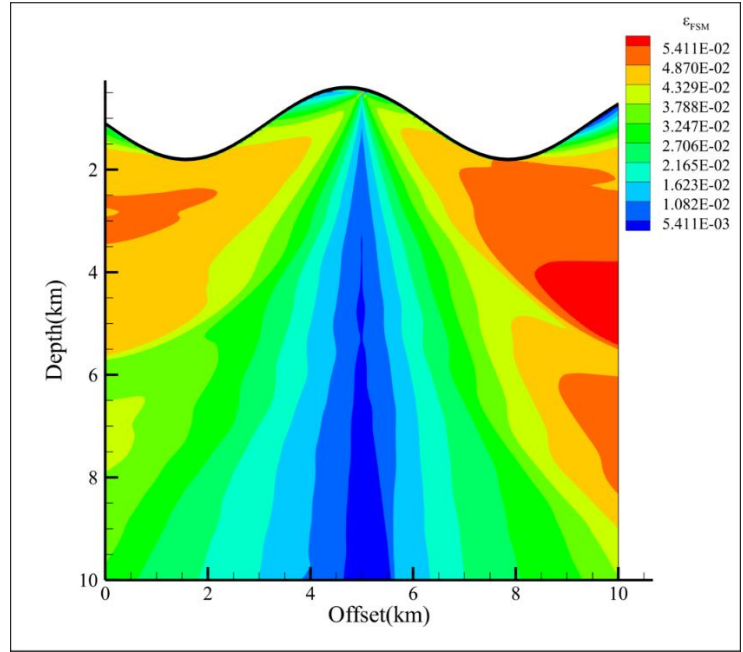

**Figure 24.** Absolute error measure in FSM against reference solution – BP TTI model

## 5.5 qSV –wave

Substituting for the local derivatives from Eq. (24) in Eq. (10) results in the discrete form of the qSV wave VTI Eikonal equation as

$$\left( \begin{array}{c} \left( \sum_{j=1}^{N_{(k)}} T_{(k)(j)} g_1^{01}(\xi_{1(k)(j)}, \xi_{2(k)(j)}) A_{(j)} \right) \cos\theta - \\ \left( \sum_{j=1}^{N_{(k)}} T_{(k)(j)} g_1^{10}(\xi_{1(k)(j)}, \xi_{2(k)(j)}) A_{(j)} \right) \sin\theta \end{array} \right)^2 - \frac{B(x_{(k)}, z_{(k)}) + \sqrt{B^2(x_{(k)}, z_{(k)}) - 4C(x_{(k)}, z_{(k)})}}{2} = 0 \quad (48)$$

in which

$$B(x_{(k)}, z_{(k)}) = \frac{1}{\upsilon_p^2} + \frac{1}{\upsilon_s^2} - 2\left(1 + \delta + (\varepsilon - \delta)\frac{\upsilon_p^2}{\upsilon_s^2}\right)\left( \begin{array}{c} \left( \sum_{j=1}^{N_{(k)}} T_{(k)(j)} g_1^{10}(\xi_{1(k)(j)}, \xi_{2(k)(j)}) A_{(j)} \right) \cos\theta \\ + \left( \sum_{j=1}^{N_{(k)}} T_{(k)(j)} g_1^{01}(\xi_{1(k)(j)}, \xi_{2(k)(j)}) A_{(j)} \right) \sin\theta \end{array} \right)^2 \quad (49)$$

and

$$C(x_{(k)}, z_{(k)}) = \left( (1+2\varepsilon) \left( \begin{array}{l} \left( \displaystyle\sum_{j=1}^{N_{(k)}} T_{(k)(j)} g_1^{10}(\xi_{1(k)(j)}, \xi_{2(k)(j)}) A_{(j)} \right) \cos\theta \\ + \left( \displaystyle\sum_{j=1}^{N_{(k)}} T_{(k)(j)} g_1^{01}(\xi_{1(k)(j)}, \xi_{2(k)(j)}) A_{(j)} \right) \sin\theta \end{array} \right)^2 - \frac{1}{\upsilon_p^2} \right)$$
$$\times \left( \left( \begin{array}{l} \left( \displaystyle\sum_{j=1}^{N_{(k)}} T_{(k)(j)} g_1^{10}(\xi_{1(k)(j)}, \xi_{2(k)(j)}) A_{(j)} \right) \cos\theta \\ + \left( \displaystyle\sum_{j=1}^{N_{(k)}} T_{(k)(j)} g_1^{01}(\xi_{1(k)(j)}, \xi_{2(k)(j)}) A_{(j)} \right) \sin\theta \end{array} \right)^2 - \frac{1}{\upsilon_s^2} \right)^2 \qquad \text{for} \quad k=1,K \qquad (50)$$

with $x \in [0, 2\text{km}]$ and $z \in [0, 2\text{km}]$. The P- and S-wave phase velocity fields with constant gradients are shown in Fig. 25 with a source at $\mathbf{x}_s = (1.0\text{km}, 1.0\text{km})$. The Thomsen anisotropy parameters have uniform values of $\varepsilon = 0.4$ and $\delta = 0.2$, and the homogeneous tilt angle is $\theta = 40°$. the domain is discretized with $K = 101 \times 101$ points. The corresponding uniform grid spacing is $\Delta = 10\,\text{m}$. The initial guess for the traveltime field is constructed with the average velocity gradients $\bar{g}_x = 0$ and $\bar{g}_z = 0.5 s^{-1}$. With $\upsilon_p(\mathbf{x}_s) = 2.5\text{km/s}$, the initial guess for traveltime becomes

$$T_{init}(\mathbf{x}) = 2\cosh^{-1}\left(1 + \frac{\left|\mathbf{x} - (1.0, 1.0)\right|^2}{20\upsilon(\mathbf{x})}\right) \qquad (51)$$

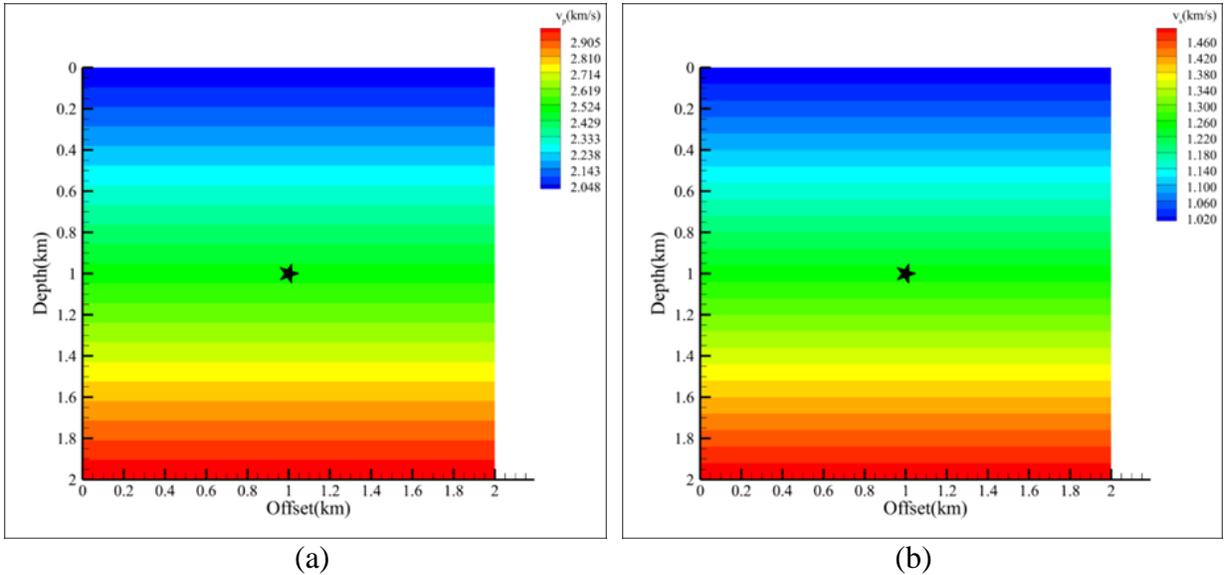

(a)                                    (b)

**Figure 25.** Velocity field in the domain with a source at $\mathbf{x}_s = (1.0, 1.0)$: a) P-, and b) S-wave

Fig. 26 shows the comparison of the PD solution with that of low-rank approximation at $t = 0.63s$ (Fomel et al. 2013). The PD solution is achieved when $\|\mathbf{F(u)}\| < 1.877 \times 10^{-13}$. The PD solution recovers the main branch of the solution due its nonlocality and numerical dissipation.

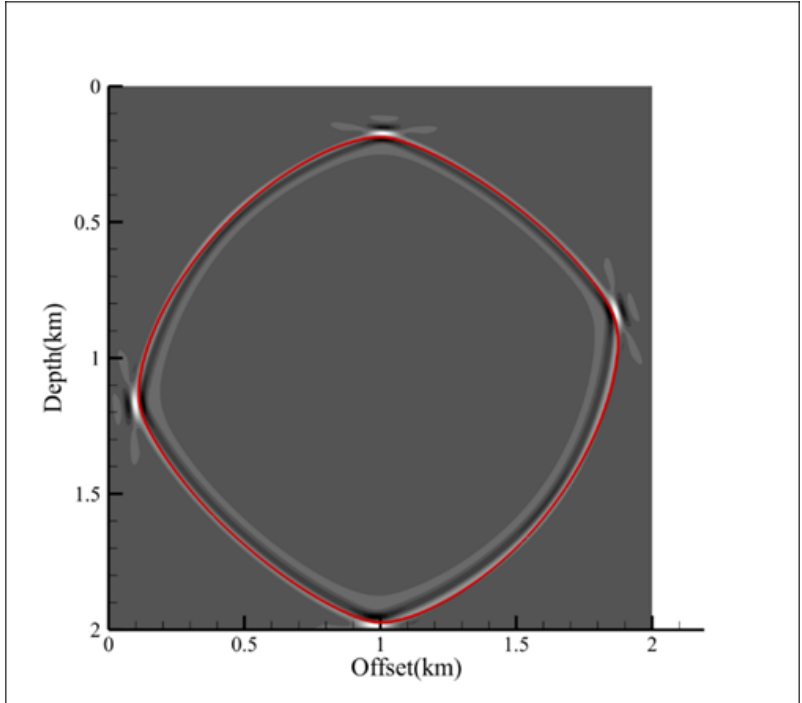

**Figure 26**. Comparison of PD prediction of traveltime contour (red) with the low-rank solution at $t = 0.63s$

## 6 CONCLUSIONS

The study presents a new approach for solving different types of Eikonal equation by using the nonlocal PDDO. The PDDO converts the local form of Eikonal equations to their nonlocal form with an internal length parameter (horizon) and a weight function with directional nonlocality. The PDDO solution handles discontinuities very well and invokes the direction of information travel in a consistent manner. The weight function controls the degree of association among points within the horizon. Also, it enables directional nonlocality based on the knowledge of characteristic directions along which information travels. Solutions are constructed in a consistent manner without any special treatments through simple discretization. The construction of initial guess for the traveltime is systematic and automatic differentiation is used to calculate the Jacobian matrix at every iteration.

The PDDO successfully provides accurate solutions to different types of challenging Eikonal equations with complex velocity fields in complex anisotropic media and with irregular topography. Unlike FSM or FMM, the PDDO does not require any modification. Tracking the gradients of traveltimes is adequate to capture both the group and phase velocity vectors; the results are stable and accurate. Although not investigated, the computational time may be more costly than the FSM and FMM. However, the PDDO is extremely suitable for parallelization, particularly on GPU (Graphics Processing Unit) using CUDA software. This feature can be explored for computational speed up especially when considering three-dimensional Eikonal equations.

## ACKNOWLEDGEMENT

This study was performed as part of the ongoing research at the MURI Center for Material Failure Prediction through Peridynamics at the University of Arizona (AFOSR Grant No. FA9550-14-1-0073).